# Multiple Cell Upset Partitioning for Simulation of Soft Error Rates in Space Systems with Error Correcting Codes

Gennady I. Zebrev, Artur M. Galimov, Liza V. Mrozovskaya,

Maxim S. Gorbunov, Konstantin A. Petrov

*Abstract* — **A self-consistent procedure for the ion-induced soft error rate calculation in space environment taking into account Error Correcting Codes is proposed. The method is based on the partitioning of the multiple cell events into groups with different multiplicities. The proposed partitioning method has been validated for the ground and on-orbit literature data.**

## I. INTRODUCTION

THE shrinking of the cell elements in modern highly-scaled memories leads to new fundamental challenges to ensure their reliable operation in harsh space environments [1]. According to [2], the Soft Error Rate (SER) per a bit is approximately proportional to the cell area $a_C$ and inversely proportional to its critical charge $\propto a_C/Q_C$. In relatively older technologies (down to ~ 65 nm), $Q_C$ decreased by roughly 30% per a generation, mainly due to capacitance lowering, whereas for the ultra-scaled devices (with the technological nodes 60-14 nm) this dependence tends to a saturation [3]. Thus, the SER per a bit is reducing during the cell shrinking, while the error rate does not diminish, but even increases at the circuit or at a system level. Error Correcting Codes (ECC) is commonly implemented to protect against the soft errors [4]. The rise of multiple cell upsets (MCU) has led to new challenges for efforts to predict the SER in the space environment. The traditional SER calculation methods and software packages such as CREME96, SPENVIS, OSOT are able to compute the MCU rate without an accounting of the ECC. The SER prediction in the devices and systems with the ECC generally requires knowledge of the multiple cell event statistical properties, which are not typically provided by the standard testing and computational procedures.

This work is aimed at development and validation of a self-consistent partitioning procedure for the multiple cell events, followed by a use of this information for calculation of SER in the systems with embedded ECC, taking into account the GCR LET spectra at a given orbit.

## II. PHYSICAL BIT-FLIP RATE SER

### A. Mathematical formalism

The single-bit upset (SBU) rate per a bit can be calculated using a functional of the mean cross section $\sigma(\Lambda)$ and the heavy ion flux LET spectrum $\phi(\Lambda)$, averaged over total solid angle

$$R_{SBU}\left[\phi(\Lambda)\right] = \int \sigma(\Lambda)\phi(\Lambda)d\Lambda. \qquad (1)$$

We have proposed and validated this approach in [5]. We emphasize that, despite the use of a linear approximation for the cross section curve in original work [5], the results of numerical integration with (1) are unaffected by a specific form interpolation of $\sigma(\Lambda)$.

The total rate of bit-flips in (1) is generally composed of the events with different multiplicities. Following the general methodology, we have shown in [6] that the total SBU cross section in (1) can be decomposed as a sum

$$\sigma(\Lambda) = \sum_{n=1} n\,\sigma_n(\Lambda) = a_C \sum_{n=1} n\, p_n(\Lambda) = a_C m(\Lambda), \qquad (2)$$

where $\sigma_n$ is a partial cross section for the events with a given multiplicity $n$, $p_n(\Lambda) = \sigma_n(\Lambda)/a_C$ is the multiplicity distribution ($n = 0$ corresponds to the event (ion hit) without effect (upset)), and $m(\Lambda) = \sigma(\Lambda)/a_C$ is the average multiplicity for a given LET. The forms of multiplicity distributions are different at different LETs and, in principle, they can be determined experimentally during the ground tests. Notice that the partial cross distributions at every LET are bound by the completeness condition $\sum_{n=0} \sigma_n(\Lambda) = a_C$.

Manuscript received September 29, 2017.

G. I. Zebrev, A. M Galimov, E. V. Mrozovskaya, M. S. Gorbunov are with Department of Micro- and Nanoelectronics of National Research Nuclear University MEPHI, 115409, Kashirskoe sh., 31, Moscow, Russia, e-mail: gizebrev@mephi.ru.

K. A. Petrov and also M. S. Gorbunov are with Scientific Research Institute of System Analysis, Moscow, Russia.

This work was supported by the Competitiveness Program of NRNU MEPHI.

Acting similarly, we have to partition the total bit-flip rate in the space environment into a number of the group events with different multiplicities. Taking into account (1) and (2), we have

$$R_{SBU}[\phi(\Lambda)] = \int_0^\infty \sigma(\Lambda)\phi(\Lambda)d\Lambda = \sum_{n=1}^\infty n R_n, \quad (3)$$

where the rate of n-folded event per a bit is defined as follows

$$R_n = a_C \int_0^\infty p_n(\Lambda)\phi(\Lambda)d\Lambda. \quad (4)$$

A sum of partial frequencies is equal to the total ion flux $\phi$ per a cell

$$\sum_{n=0} R_n = a_C \int \phi(\Lambda)d\Lambda = a_C \phi \equiv R_{tot}. \quad (5)$$

Then we can define an effective cross section averaged both over the LET spectrum of a given orbit and over the multiplicity distribution

$$\sigma_{eff} = \frac{\int \sigma(\Lambda)\phi(\Lambda)d\Lambda}{\int \phi(\Lambda)d\Lambda} = a_C \frac{\sum_{n=0} n R_n}{\sum_{n=0} R_n} \quad (6)$$

This is a figure-of-merit of a given circuit, which completely characterizes the SBU SER at a given orbit

$$R_{SBU} = \sigma_{eff}\phi. \quad (7)$$

For numerical estimation, we will assume that the low limit for integration over LETs equals 0.1 MeV-cm$^2$/mg. Given that LET spectra at different orbits are often different from each other approximately by a constant, the parameter $\sigma_{eff}$ can be considered as a very useful and informative figure-of-merit for the soft error rates of the unscreened circuits.

### B. Poisson conjecture

So far our theoretical scheme had a general form and did not use the explicit form of the multiplicity distribution. There are physical reasons to believe that the multiplicity distribution $p_n$ in the MCU effects are often close to the Poisson distribution [6]. The direct experimental validation of this assumption is presented in the Appendix. Defining the Poisson distribution as follows

$$p_n(\Lambda) = \frac{m(\Lambda)^n}{n!}e^{-m(\Lambda)}, \quad (8)$$

we get the self-consistent value of the average cross section (2) in a consistent way at each LET point.

Then, the partial rate of events with a multiplicity $n$ per a bit is given as follows

$$R_n = a_C \int_0^\infty p_n(\Lambda)\phi(\Lambda)d\Lambda = a_C \int_0^\infty \frac{m(\Lambda)^n}{n!}e^{-m(\Lambda)}\phi(\Lambda)d\Lambda. \quad (9)$$

Particularly, $R_0$ is a frequency of the "zero" events (i.e., the ion strikes without any upsets), which is given by

$$R_0 = a_C \int e^{-m(\Lambda)}\phi(\Lambda)d\Lambda; \quad (10)$$

$R_1$ is a frequency of the single bit-flips and so on

$$R_1 = a_C \int m(\Lambda)e^{-m(\Lambda)}\phi(\Lambda)d\Lambda. \quad (11)$$

For rad-hard RHBD devices in which in the entire or in a greater part of the LET range $m(\Lambda) \ll 1$, we have $R_{tot} \cong R_0 + R_1$ and $R_{SBU} \cong R_1 \cong a_C \int m(\Lambda)\phi(\Lambda)d\Lambda$.

### C. Linear approximation

For simplicity, we will use in this work the linear approximation for the cross section dependence $\sigma(\Lambda) \cong K_d(\Lambda - \Lambda_C)$ which is well adopted for highly scaled circuits with low critical LET $\Lambda_C$ [2, 5, 6]. We have in this case

$$m(\Lambda) = \sigma(\Lambda)/a_C \cong K_d(\Lambda - \Lambda_C)/a_C. \quad (12)$$

This specific approach is not, however, critical, since any approximation for $\sigma(\Lambda)$ (including a formal numerical interpolation) can be used for a numerical procedure of partitioning of events into groups with different multiplicities.

## III. PARTITIONING EXAMPLES AND APPLICATIONS

We have compared in [5] the on-board data and the total SER simulated with (1). Here we present the results of partitioning of the total SER for some data.

### A. Proba II data

The calculated partial frequencies for 4 Mbit 0.25 μm devices at PROBA-II mission are shown in Fig. 1 [7].

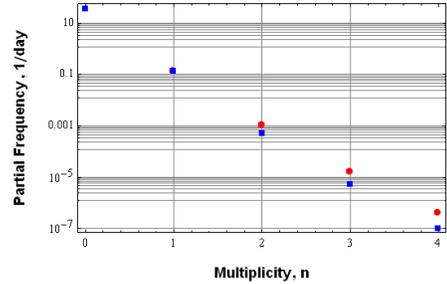

Fig. 1. Partial SBU frequency $n \times R_n$ (red circles) and the partial n-fold event frequency $R_n$ (blue squares), calculated as functions of multiplicity $n$ for the Atmel AT68166 16 Mbit SRAMs [7]. Total SER 0.133 (vs in-flight 0.138) errors/day, effective cross section $\sigma_{eff} = 5.4 \times 10^{-11}$ cm$^2$.

The effective cross section per a bit $\sigma_{eff}$ was also estimated for the Proba II orbit.

## B. SAC-C Mission Data

Figures 2 and 3 show the partitioning results for the two types of 4 Mbit SRAMs aboard the SAC-C Mission [8, 9].

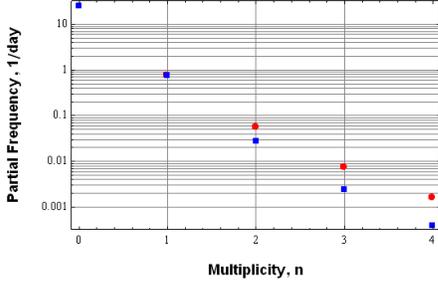

Fig. 2. Partial SBU frequency (red circles) and partial n-fold event frequency $R_n$ (blue squares), calculated as functions of multiplicity $n$ for the 4 Mbit SRAMs KM684000. The calculated total SER is 0.85 errors/day, effective cross section $\sigma_{eff} = 3.4 \times 10^{-10}$ cm$^2$.

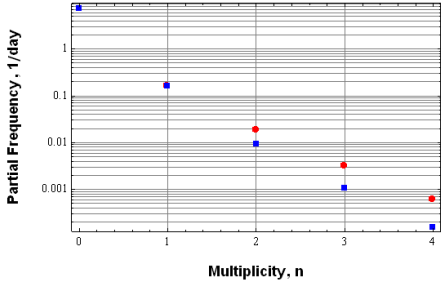

Fig. 3. Partial SBU frequency (red circles) and partial n-fold event frequency $R_n$ (blue squares), calculated as functions of multiplicity $n$ for the 4 Mbit SRAMs HM628512. Calculated total SER is 0.19 errors/day, effective cross section $\sigma_{eff} = 7.5 \times 10^{-11}$ cm$^2$.

The cell area in all calculations was assumed to be $a_C = 20$ μm$^2$ (an approximate estimation for 0.25 μm technology node).

We have compared the simulation results for SAC-C mission's SRAMs with available in-flight data [9] (see Fig 4. and 5).

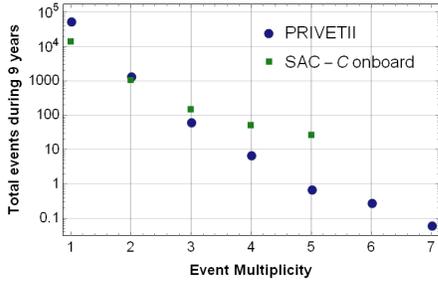

Fig. 4. Simulated (circles) and in-flight (squares) partial n-fold event rate for KM684000.

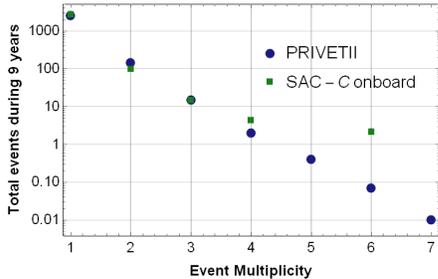

Fig. 5. Simulated (circles) and in-flight (squares) partial n-fold event rate for HM628512.

This comparison demonstrates excellent agreements between experiment and simulation for two types of devices at least up to $n = 3$ and good and satisfactory agreements for $n = 4$.

## C. SER estimation at a system level

The distribution of the event rates over multiplicities $R_n[\phi(\Lambda)]$ in a given device or a system carries exclusively important information about their vulnerabilities to SEE in a given space environment. Accurate knowledge of the MCU distributions is also crucial for determining the interleaving design rules. In practice, all the correcting codes are unable to systematically fix the single events with a sufficiently large multiplicity without a significant increase in the hardware and time costs.

Basically, any ECC algorithm can provide the error correction and/or detection with a certain probability, which depends on the event multiplicity. If we have the vector $V_n$, characterizing the probability of the error missing for the n-folded event [10], [11], the system error rate at a given orbit can be estimated as follows

$$R_{syst} = \sum_{n=1} V_n R_n, \quad (13)$$

or,

$$R_{syst} = \sum_{n=1} n V_n R_n, \quad (14)$$

depending on the task context.

## D. Scrubbing efficiency for a simple SEC-DED procedure

The system ECC word error rate can be calculated as follows

$$R_{syst} = \beta M R_w(n \geq 1) + M R_w(n \geq 2) \quad (15)$$

where $N_W$ is the total number of words, $N_{BW}$ is the number of bits in the word, $M = N_W N_{BW}$ is the total memory capacity, $R_w(n \geq 1)$ (or $R_w(n \geq 2)$) is the Multiple Bit Error (MBU) rate (i. e., the MCU rate in a single word) with multiplicity $n \geq 1$ (or $n \geq 2$). The dimensionless parameter $\beta$ is defined as follows

$$\beta = \frac{1}{2}(N_{BW} - 1)(R_1 t_S), \quad (16)$$

where $t_S$ is the scrubbing time interval, $R_1$ is the rate of the single bit-flips. The basic formula (15) was obtained in a standard approximation, assuming a reasonable condition $\beta \ll 1$. The first term in (15) is the well-known Saleh-Edmonds relation [12], modified for a general case of MCU impact. This term describes an action of a simple SEC-DED algorithm for correction of the two successive errors, the first of which is a single bit-flip while the second is any type of soft errors. The second term in (15) corresponds to Multiple Bit Error (MBU) rate. Generally, $R_w(n \geq 1) > R_w(n \geq 2)$, but aggressive error correction via $t_S$ (and $\beta$) reducing is capable to significantly suppress the first term in (15).

Let us consider a very simple illustrative example. The following reasonable approximations could be done for the $R_w$

$$R_w(n \geq 1) \cong R_1 + (1/2) R_2, \quad (17)$$
$$R_w(n \geq 2) \cong (1/2) R_2, \quad (18)$$

where all the event rates with $n \geq 3$ are neglected because of their assumed smallness. A factor 1/2 appeared in (17) and (18) due to an assumption that only a half of the two-fold MCUs locate in the same word. Then, using (17), we get

$$V_n \cong \{\beta, (1/2)(1+\beta), 1, 1, ...\}. \quad (19)$$

This means, that a 'perfect' SEC-DED scrubbing procedure ($\beta = 0$) fully suppresses all the single bit-flips, and approximately a half of the double bit-flips, missing all the events with $n \geq 3$. Such, or a similar method could be used to evaluate the trade-offs between vulnerability to SEE and performance of the devices and systems at the early stages of design [13].

## IV. CONCLUSION

We have proposed and validated a simple self-consistent technique to partition the MCU events into groups with different multiplicities based only on the standard mean cross section vs LET testing data. The results of such partitioning can be used for the error rate calculation in devices and systems provided by ECC.

## APPENDICES

### A. Reduced Poisson distribution

To validate the Poisson distribution conjecture we have to define another (reduced) Poisson distribution, inferred from original by exclusion of 'zero' events

$$\tilde{p}_n = \frac{p_n}{1-p_0} = \frac{m^n}{n!} \frac{e^{-m}}{1-e^{-m}} = \frac{m^n}{n!} \frac{1}{e^m - 1}, \quad n = 1, 2, 3 ... \quad (A1)$$

Both distributions are normalized

$$\sum_{n=0}^{\infty} p_n(\Lambda) = \sum_{n=1}^{\infty} \tilde{p}_n(\Lambda) = 1. \quad (A2)$$

The reduced Poisson distribution does not contain information about the number of 'zero events' (ion strikes without any observed effects) and it represents the distribution of the events only with multiplicities $n \geq 1$. Both distributions are shown in Fig. 6.

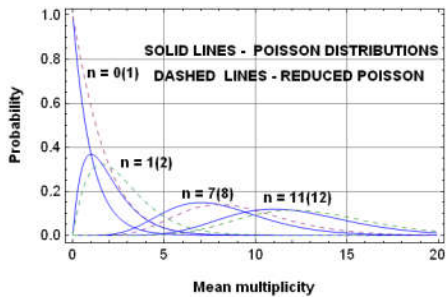

Fig. 6. The probability functions the Poisson $p_n$ (solid lines) and the reduced Poisson $\tilde{p}_n$ (dashed lines) distributions at for different multiplicities.

An extremely low mean multiplicity (or, the same, mean cross section) corresponds to close to unity probabilities of 'zero events' and single-bit effects. Notice that the reduced distribution is suitable only for comparison with experimental data since most of the literature data correspond to such type of distribution.

### B. Empirical validation of Poisson conjecture

Figures 7 and 8 show comparison of simulated with (A1) probabilities and distributions on LET. The empirical data for multiplicity distributions in commercial SRAMs at different LETs are taken from [14, 15].

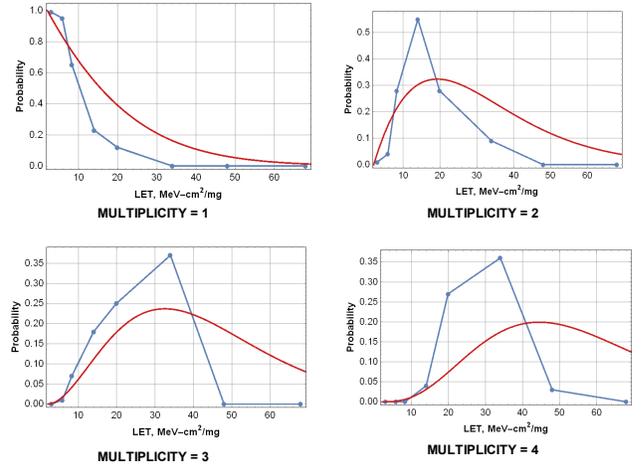

Fig. 7. Comparison of the reduced Poisson distribution results (smooth lines) calculated from the cross section vs LET dependence ($K_d = 0.48 \times 10^{-9}$ mg/MeV, $\Lambda_C = 2$ MeV-cm$^2$/mg, $a_C = 0.52$ μm$^2$) with the experimental distributions (joined circles), taken from Fig.5. [12].

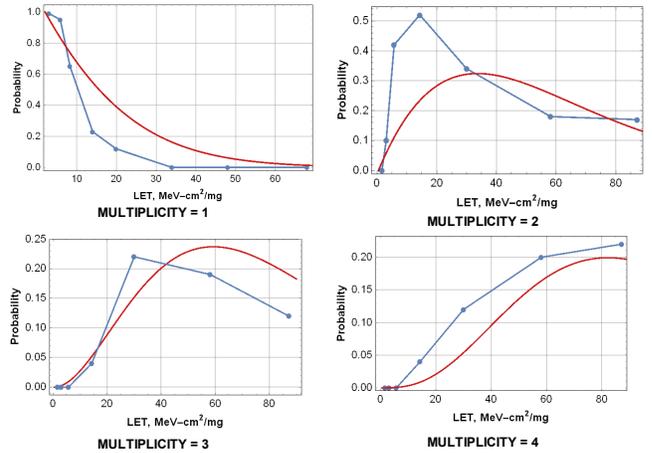

(c) Data adopted from [Lawrence-2008, Tab. 2]

Fig. 8. Comparison of the the reduced Poisson distribution (smooth lines) calculated from the cross section vs LET dependence ($K_d = 0.40 \times 10^{-9}$ mg/MeV, $\Lambda_C = 0.5$ Mev-cm$^2$/mg, $a_C = 1$ μm$^2$) with the experimental distributions (joined circles), taken from [13].

As can be seen from the figures above the Poisson formalism is quite satisfactorily simulates the MCU distribution on LET.